\documentclass[twocolumn,nofootinbib,superscriptaddress]{revtex4-1}

\pdfoutput=1

\usepackage{slashed}

\usepackage[normalem]{ulem}

\usepackage{color}

\usepackage{epsfig}
\usepackage{epstopdf}
\usepackage{epsf}
\input{epsf.sty}
\usepackage{graphicx,amsmath,amssymb}
\usepackage{epsfig}
\allowdisplaybreaks

\def\ei{\end{itemize}}
\def\be{\begin{equation}}
\def\ee{\end{equation}}
\newcommand{\bea}{\begin{eqnarray}}
\newcommand{\eea}{\end{eqnarray}}

\usepackage{bbm,bm,amsmath,amssymb}

\usepackage{hyperref}

\def\Kahler{K\"{a}hler~}

\begin{document}

\begin{flushleft}
 DESY 16-172 \\ \textcolor{white}{blank space}
\end{flushleft}
\vspace{0mm}

\title{Inflation from Nilpotent K\"ahler Corrections} 

\author{Evan McDonough}
\email{evanmc@physics.mcgill.ca}
\affiliation{Ernest Rutherford Physics Building, McGill University,\\ 
3600 University Street, Montr{\'e}al QC, Canada H3A 2T8\\}
\author{Marco Scalisi}
\email{marco.scalisi@desy.de}
\affiliation{Deutsches Elektronen-Synchrotron, DESY, \\ Notkestra{\ss}e 85, 22607 Hamburg, Germany}

 \begin{abstract}

We develop a new class of supergravity cosmological models where inflation is induced by terms in the K\"ahler potential which mix a nilpotent superfield $S$ with a chiral sector $\Phi$. As the new terms are non-(anti)holomorphic, and hence cannot be removed by a K\"ahler transformation, these models are intrinsically K\"ahler potential driven. Such terms could arise for example due to a backreaction of an anti-D3 brane on the string theory bulk geometry. We show that this mechanism is very general and allows for a unified description of inflation and dark energy, with controllable SUSY breaking at the vacuum. When the internal geometry of the bulk field is hyperbolic, we prove that small perturbative K\"ahler corrections naturally lead to $\alpha$-attractor behaviour, with inflationary predictions in excellent agreement with the latest Planck data.

 \end{abstract}

\maketitle


\section{Introduction}

One of the key goals of modern theoretical physics is to find a UV complete description of our Universe, unifying particle physics with cosmology at both early and late times. There has recently been significant advancements towards this goal: It has indeed been realized that obtaining a pure {\it acceleration} phase, in the context of supergravity and/or string theory, often involves the appearance of a \emph{nilpotent superfield} $S$. The latter is constrained by the condition \cite{nilpotent}
\be\label{nilpotent}
S^2=0\,,
\ee
which implies the absence of scalar degrees of freedom\footnote{After writing $S$ in terms of the superspace coordinates, one can easily check that the scalar part is replaced by a bilinear fermion. This simply reflects that, in this case, supersymmetry is non-linearly realized (see also the recent investigations \cite{Nonlinear}). One can then recover the original Volkov-Akulov action \cite{VA}.}. This fact has turned out to be beneficial for cosmological applications: one can indeed show that coupling a nilpotent field to an inflationary sector generally simplifies the overall dynamics and allows for a unified description of inflation and dark energy \cite{NilpotentInflation,NilpotentInflation2}.

The initial investigations of the nilpotent chiral multiplet in the context of global supersymmetry \cite{nilpotent} (see also \cite{SUSYnilpotent}), have been by now extended to the regime when this symmetry becomes local \cite{dS_SUGRA,dS_SUGRA2}. The recent discovery of `de Sitter supergravity' \cite{dS_SUGRA}, nearly forty years after the advent of AdS supergravity \cite{Townsend:1977qa}, has marked a serious development in the field. Coupling a nilpotent multiplet to supergravity indeed gives rise to a pure de Sitter phase, with no scalar fields involved.

Some of the most interesting aspects have however emerged in the context of string theory. It was realized in \cite{antiD3Goldstino} that the four dimensional description of anti-D3 branes in an $\mathcal{N}=1$ flux background, famously used by `KKLT' to construct de Sitter vacua in string theory \cite{KKLT}, is indeed a supergravity theory of a nilpotent superfield, wherein supersymmetry is non-linearly realized. This demonstrated that supersymmetry breaking by $\overline{\rm D3}$'s is spontaneous rather than explicit, providing strong evidence for the compatibility of `uplifting' via $\overline{\rm D3}$'s and moduli stabilization via various contributions to the superpotential.

The string theory origins of constrained superfields, and connections to D-brane physics, were then further worked out in  \cite{String_Theory_Goldstinos,Dasgupta:2016prs, WraseRecent1, WraseRecent2, Bandos:2016xyu}. The fermions arising when one or more anti-branes, placed in certain geometries, break supersymmetry spontaneously (see e.g. \cite{Dasgupta:2016prs}) can often be packaged into constrained superfields. An example is provided by the recent works \cite{WraseRecent2}, where both a nilpotent superfield and an `orthogonal nilpotent superfield', as used to construct inflationary models in \cite{orthogonal}, emerge when an $\overline{\rm D3}$ is placed on intersecting O7-planes. 
It thus appears that constrained superfields are ubiquitous in string theory, and not just the simple example of a single nilpotent superfield originally studied in \cite{antiD3Goldstino}.

In terms of physical applications, the nilpotent superfield $S$ has proven to be an optimal tool in the construction of cosmological scenarios \cite{NilpotentInflation,NilpotentInflation2} (see also the recent work \cite{Linde:2016bcz}). On the one hand, it allows to easily uplift models of inflation in supergravity, analogous to what happens in string scenarios with the addition of one or many $\overline{\rm D3}$'s (see e.g. \cite{Burgess}). On the other hand, it generically ameliorates the stability properties of the model and yields better control over the phenomenology. Then, by means of an inflaton sector $\Phi$ and a nilpotent one $S$, it is possible to obtain a comprehensive physical framework which describes the primordial expansion of the Universe together with controllable level of dark energy and SUSY breaking.

However, in general there is no reason to expect the nilpotent field to be totally decoupled from the inflationary physics. In the context of string theory, this becomes a question of backreaction of the $\overline{\rm D3}$ on the bulk geometry, which would manifest itself in the $d=4$ supergravity theory as couplings between the nilpotent superfield and the bulk moduli. Typically, it is the task of model builders to argue that such corrections do not affect the dynamics of the model under consideration, and in particular, that the corrections to the K\"{a}hler potential are suppressed and do not lead to an  $\eta$-problem \cite{Copeland:1994vg,Baumann:2014nda}.  The effect on inflation due to non-negligible corrections to the K\"{a}hler potential has been considered in e.g. \cite{LargeCorrections}. Computing these corrections explicitly in a concrete string compactification setting is a notoriously difficult task and has been done in only a small number of cases, see e.g. \cite{StringyCorrections}.

In this letter, we take the opposite approach. We will show that inflation can be driven by corrections to the K\"{a}hler potential of the form
\begin{equation}
\label{mixing}
\delta K = \delta K (\Phi, \bar{\Phi}, S , \bar S),
\end{equation}
which mix the nilpotent superfield $S$ with a bulk modulus $\Phi$, even in the \emph{absence of a superpotential} for $\Phi$. This is similar in spirit to `K\"{a}hler uplifting' \cite{khleruplifting}, where the term responsible for the uplift to dS is an $\alpha'$ correction to the K\"{a}hler potential.

We will prove that this procedure provides a unified description of early and late time cosmology, where, at least qualitatively, the inflationary dynamics is due to the backreaction of the $\overline{\rm D3}$ on the bulk manifold and/or fluxes.

We will focus our investigation on the case of both {\it flat} and {\it hyperbolic} K\"ahler geometry, the latter typically arising from string theory compactifications. Therefore, in Sec.~\ref{SEC.FLAT}, we will thoroughly study the effects of the possible \Kahler corrections to a flat \Kahler geometry, providing all the relevant formulas and results. In Sec.~\ref{SEC.HYP}, we will discuss the hyperbolic case highlighting the differences and similarities with the previous flat case. Interestingly, when the internal manifold is hyperbolic, we will prove that small perturbative \Kahler corrections automatically lead to {\it $\alpha$-attractor} \cite{alphaattractors} behaviour with cosmological predictions in great agreement with the latest observational data \cite{CosmologicalData}. We will conclude in Sec.~\ref{SEC.Discussion} with a summary of the main findings and perspectives for future directions. Throughout the paper, we will work in reduced Planck mass units ($M_{Pl}=1$).

\vspace{-3mm}
\section{Inflation From K\"{a}hler Corrections to Flat Geometry}\label{SEC.FLAT}

In this section, we would like to show how inflation can arise simply from corrections to a \Kahler potential with {\it zero curvature}, while keeping the superpotential independent of the inflaton superfield. 

Our starting point is the simplest realization of de Sitter phase in supergravity. This can be encoded in the following set of K\"ahler and superpotential \cite{NilpotentInflation, dS_SUGRA}:
\be\label{KWdS}
K= S \bar{S} \,, \qquad W= W_0 + M S\,,
\ee
where $W_0$ is the flux induced superpotential \cite{Giddings:2001yu}
and $M$ parametrizes the contribution of the $\overline{\rm D3}$. Then, the resulting scalar potential is a cosmological constant of the form
\be\label{potdS}
V= M^2 - 3 W_0^2\,.
\ee
Note that one can obtain the latter Eq.~\eqref{potdS} by employing the usual formula for the scalar potential in supergravity and declaring that $S=0$ (since the bilinear fermion, replacing the scalar part of $S$, cannot get any vev). The constant phase described by Eq.~\eqref{potdS} is then the result of the delicate balance between the scale of spontaneous supersymmetry breaking of $S$
\be
D_S W= M\,,
\ee
and the gravitino mass
\be\label{gravitino}
m_{3/2}=W_0\,.
\ee

Now, let us just extend the internal K\"ahler geometry with a chiral multiplet $\Phi$ which eventually will play the role of the inflaton. In a string theory interpretation, this framework would describe a $\overline{\rm D3}$ brane (encoded in $S$) and a bulk geometry and/or fluxes (encoded in $\Phi$). Specifically, for the sake of simplicity,  we choose a flat shift-symmetric K\"ahler function for $\Phi$ and then have
\be\label{KWnil}
K= -\tfrac{1}{2}\left(\Phi-\bar{\Phi}\right)^2+S\bar{S}\,,\qquad W= W_0+MS\,.
\ee
The latter setting still provides the same constant value \eqref{potdS}, along the real axis Im$\Phi=0$. This is obviously a flat direction as both $K$ and $W$ do not depend on Re$\Phi$. On the other hand, the orthogonal field Im$\Phi$ has a positive mass when $|M|>|W_0|$. However, this direction turns out to be not suitable for inflation, as it is too steep (due to the typical exponential dependence $e^K$ in the scalar potential).

In order to produce inflationary dynamics, one must break the shift symmetry of $K$. Traditionally, this has been done by introducing a $\Phi$-dependence in the superpotential (see for example the pioneering work \cite{Kawasaki:2000yn} and the subsequent developments \cite{mixingTerms}). In the context where $S$ is nilpotent, this approach has been put forward by \cite{NilpotentInflation}. The basic idea is to promote $W_0$ and $M$ in Eq.~\eqref{KWnil} to functions of the field $\Phi$. This breaks the shift symmetry along the real axis of $\Phi$ and perturbs the original flat direction creating an inflationary slope.

In this letter, we intend to explore an alternative possibility to induce inflation: while keeping a $\Phi$-independent superpotential, we can add terms to the K\"ahler potential which mix the two sectors, break the original shift-symmetry and encode the interaction between the antibrane and the bulk modulus. 

In full generality, the only possible allowed corrections are either {\it bilinear} or {\it linear} in $S$ and $\bar{S}$, such as
\be\label{corrections}
\delta K = f(\Phi,\bar{\Phi}) S \bar{S} + g(\Phi,\bar{\Phi}) S + \bar{g}(\Phi,\bar{\Phi})\bar{S}\,,
\ee
 where $f$ and $g$ are arbitrary functions of their arguments, whose non-zero values can break the shift symmetry for\footnote{In the context where $S$ is an unconstrained chiral multiplet, the works \cite{mixingTerms} already considered bilinear couplings. However, these were taken to be independent on Re$\Phi$, thus not affecting the form of the inflationary potential.} Re$\Phi$. Note that higher order terms in $S$ are forbidden since this field is nilpotent and  Eq.~\eqref{nilpotent} holds. In addition, the above couplings are in general non-(anti)holomorphic, and so cannot be gauged away by a K\"{a}hler transformation (whereas this is possible when $S$ satisfies also an orthogonal nilpotency constraint \cite{orthogonal}).
 
 It is interesting to notice that the corrections \eqref{corrections} will affect the form of the \Kahler metric such as
\be
K_{I \overline{J}} = 
 \begin{pmatrix}
  1 & \partial_{\bar{\Phi}} \bar{g} \\
  \partial_{{\Phi}} {g} & 1+f
 \end{pmatrix}\,,
\ee
thus inducing non-zero off-diagonal terms and modifying the originally canonical $K_{S\bar{S}}$.

However, this turns out not to be an issue for the cosmological dynamics of the model as the field $S$ is nilpotent (fermion interactions are subdominant during inflation) and the only scalars involved are the real and imaginary components of $\Phi$. The off-diagonal terms may have some relevant consequences for the post-inflationary evolution, as we comment in the concluding section of this paper.

 In the following, we analyse the effects of the bilinear and linear nilpotent corrections separately.

\subsection{Bilinear nilpotent corrections}\label{SUBSECbilinear}

Let us focus on the effects of the sole bilinear corrections while keeping $g=0$. The model is then characterized by the same $\Phi$-independent superpotential given in Eq.~\eqref{KWnil} and a \Kahler potential such as
\be\label{BilinearCorrectionsK}
K= -\tfrac{1}{2}\left(\Phi-\bar{\Phi}\right)^2+\left[1+f(\Phi,\bar{\Phi}) \right]S \bar{S}\,.
\ee
This class of couplings is well motivated from string theory as the \Kahler potential for D-brane matter fields generically appears as a bilinear combination of the fields and their complex conjugate.

This model still allows for an extremum along $\Phi=\bar{\Phi}$ (i.e. Im$\Phi=0$) if the function $f$ satisfies
\be\label{conditionextremumBI}
\partial_\Phi f(\Phi,\bar{\Phi})|_{\Phi=\bar{\Phi}}=\partial_{\bar{\Phi}} f(\Phi,\bar{\Phi})|_{\Phi=\bar{\Phi}}\,.
\ee
A sufficient condition for Eq.~\eqref{conditionextremumBI} to be valid is that $f$ is symmetric under\footnote{This is analogous to the reality property imposed on the holomorphic function $f(\Phi)$ in the superpotential of the models \cite{mixingTerms} in order to assure consistent truncation along the real direction.} $\Phi \leftrightarrow \bar{\Phi}$. Then, typical corrections are the ones depending on $(\Phi+\bar{\Phi})$ or  $\Phi\bar{\Phi}$. In addition, in order for Im$\Phi=0$ to be a consistent truncation, one must ensure positive mass of the orthogonal direction (we discuss this later).

Supersymmetry is spontaneously broken just in the $S$-direction as the F-terms are equal to
\be \label{Fterms}
D_{\Phi} W=0\,, \qquad D_{S} W= M\,,
\ee
for any value of $\Phi$ and then for the entire cosmological evolution. Note the difference with respect to the previously developed nilpotent cosmological models \cite{NilpotentInflation,NilpotentInflation2}, which yield a positive potential thanks to the supersymmetry breaking along both directions.

One can simplify the following discussion by defining the function
\be\label{functionF}
 F(\Phi)\equiv\frac{1}{1+f(\Phi,\Phi)}\,,
\ee
along the extremum $\Phi=\bar{\Phi}$.

The combined effects of the SUSY breaking in $S$ and of the non-zero \Kahler correction generates a scalar potential for $\Phi$, along the real axis, given by
\be\label{potbilinear}
V(\Phi)= -3W_0^2 + M^2 F(\Phi)\,,
\ee
which clearly allows for arbitrary inflation and a residual cosmological constant (CC). At the minimum of the potential (which is placed at $\Phi=0$, provided $F'(0)=0$), we have indeed 
\be
\label{CC}
\Lambda=-3W_0^2 + M^2 F(0)\,.
\ee

The cosmological constant and amplitude of the inflationary potential are thus determined in terms of the same underlying parameters. The cosmological constant is constrained to be very small by late-time cosmology, while the size of the inflationary potential is fixed (albeit in a model-dependent way) by the amplitude of the curvature perturbations in the cosmic microwave background (CMB). 

Note that, within this framework, a large value of $M$ does not necessarily correspond to a very high gravitino mass, which is still equal to $W_0$, as in Eq.~\eqref{gravitino}. At the vacuum, the SUSY breaking scale is indeed given by
\be
K^{S\bar{S}}|D_S W|^2= M^2 F(0)\,,
\ee
where the \Kahler metric term $K^{S\bar{S}}$ is non-canonical, unlike the dS model defined by Eq.~\eqref{KWdS} and the models of \cite{NilpotentInflation,NilpotentInflation2} (in these cases the almost vanishing CC forces $M$ and $m_{3/2}$ to be of the same order). A small fine-tuned value of $F(0)$ can still allow for a desirable low gravitino mass (e.g. order TeV) and a negligible cosmological constant (in the spirit of the string theory landscape). Nevertheless,  the latter case (small $F(0)$ and $m_{3/2}$) implies a large \Kahler correction $f$ at the minimum and thus a considerable deviation of $K$ from its canonical form Eq.~\eqref{KWnil}.

The regime of small \Kahler corrections $|f|\ll1$ corresponds instead to an $F$ of order unity. In this case, Eq.~\eqref{potbilinear}  implies that the parameter $M$ must be of the same of order of the Hubble scale of inflation $H$ or even higher, such as
\be
M\geq H\,.
\ee
This holds during the whole cosmological evolution until the minimum of the potential, since $M$ is a constant. In this limit, the scalar potential can be indeed expanded as 
\be\label{potbilinearsmall}
V= (M^2 - 3 W_0^2) - M^2 f + \mathcal{O}(f^2)\,,
\ee
which makes once more explicit what we just said about the magnitude of $M$. Note that, in this regime, the CC is given again as compensation between $M$ and the gravitino mass, as in Eq.\eqref{potdS}. Therefore, the latter $m_{3/2}=W_0$ is necessarily large.

To summarize, bilinear nilpotent corrections to a flat \Kahler potential, such as the ones of Eq.~\eqref{BilinearCorrectionsK}, can account for both inflation and dark energy. Both acceleration phases are solely due to spontaneous SUSY breaking of the nilpotent field $S$. The non-trivial structure of the K\"{a}hler correction still allows to have great control over the phenomenology of the cosmological model, with tunable level of the CC and the scale of SUSY breaking.

\subsubsection*{Stability}

It is important that we check the stabilization of Im$\Phi$. The mass of Im$\Phi$ is given by,
\be
m_{{\rm Im}\Phi} ^2 = - 4 W_0 ^2 +   4 M^2 F(\Phi)\,.
\ee
The mass at late times, at the minimum of the potential, is given by
\be
\label{massa}
m_{{\rm Im}\Phi}  ^2 = 8 W_0 ^2 + 4 \Lambda\,,
\ee
where we have used Eq.\eqref{CC} to relate $M$ and $W_0$.

The mass during inflation, expressed as a ratio to the Hubble constant $H ^2 \sim \frac{1}{3} V$, is given by
\be
\label{massaInflation}
\frac{m_{{\rm Im}\Phi}  ^2}{H^2} = 12 + \frac{24 W_0 ^2}{M^2 F(\Phi) -  3 W_0 ^2 } .
\ee
Both the above terms are positive, and the first term is dominant. The above ratio is large and Im$\Phi$ is effectively stabilized during inflation, regardless of the precise details of $F$, $W_0$ or $M$.

\subsubsection*{Example: quadratic inflation }

As a concrete example, let us consider the classic model of quadratic inflation. In the following, we  explicitly construct this model in our framework. We do so in two ways, which have low and high scale supersymmetry breaking respectively.

First consider the following K\"{a}hler potential,
\be
\label{m2phi2lowscale}
K= -\tfrac{1}{2}\left(\Phi-\bar{\Phi}\right)^2+ \frac{M^2 }{M^2 +m^2 \Phi \overline{\Phi} } S\bar{S} ,
\ee
which corresponds to the choice
\begin{equation}\label{Fquadraticinflation}
F(\Phi) = \frac{m^2}{M^2} \Phi^2 + 1   .
\end{equation}
This gives the scalar potential
\begin{equation} \label{quadraticpotential}
V =\left(  M^2 - 3 W_0 ^2 \right) + \frac{1}{2} m^2 \varphi^2\,,
\end{equation}
with $\varphi=\sqrt{2}$Re$\Phi$.
The normalization of the inflationary potential depends only on $m$, and hence the only constraint on $M$ and $W_0$ comes from the condition that $\Lambda$ be small. Thus this model allows for low-scale supersymmetry breaking and a small gravitino mass. In this case, the magnitude of $f = (1/F) - 1$ is necessarily very large during inflation and hence the model is a large deviation from a canonical K\"ahler potential.

We can also construct this model as a small perturbative correction away from a flat K\"{a}hler potential. Consider the following K\"{a}hler potential:
\be
\label{m2phi2perturbative}
K= -\tfrac{1}{2}\left(\Phi-\bar{\Phi}\right)^2+S\bar{S} - \, \frac{m^2}{2 M^2} \Phi \bar{\Phi} \cdot S\bar{S}, \\
\ee
which corresponds to the choice of $f(\Phi,\bar{\Phi})$,
\begin{equation}
f(\Phi,\bar{\Phi}) = - \frac{m^2}{2 M^2} \Phi \bar{\Phi} .
\end{equation}
If we impose the condition that $|f| \ll 1 $, so as to be a small correction to a flat K\"{a}hler potential, this again gives the same quadratic potential \eqref{quadraticpotential}.

The normalization $V\sim 10^{-10}$ when the CMB pivot scale exits the horizon during inflation (see e.g. \cite{Creminelli:2014fca}), along with the the condition $|f| \ll 1$, then imposes a condition on $M$:
\be
M^2 \gg 10^{-10} .
\ee
Since $|f| \ll 1$ implies $F \sim 1$, this corresponds to high-scale supersymmetry breaking and (due to Eq.~\eqref{quadraticpotential} and the smallness of the CC) also to a very large gravitino mass $m_{3/2}$.

\subsection{Linear nilpotent corrections}\label{SUBSEClinear}
Let us now consider terms in the K\"{a}hler potential which are linear in $S$ and $\bar{S}$ as given by Eq.~\eqref{corrections}, while neglecting the effects of the bilinear correction ($f=0$). The most general form of this correction is
\be
 \delta K =  g(\Phi,\bar{\Phi}) S + \bar{g}(\Phi,\bar{\Phi})\bar{S}\,.
\ee
If we make the simplifying assumption that $g$ is purely real ($g=\bar{g}$), then this is a coupling of $\Phi$ to Re$S$, while if $g$ is purely imaginary ($g=-\bar{g}$) then the coupling is to Im$S$.

In the former case, the model is characterized by a \Kahler potential such as
\be\label{BilinearCorrectionsK}
K= -\tfrac{1}{2}\left(\Phi-\bar{\Phi}\right)^2+S \bar{S} + g(\Phi,\bar{\Phi}) (S + \bar{S})\,.
\ee
and the same $\Phi$-independent $W$ as in Eq.~\eqref{KWnil}. Similar to the previous case of Sec.~\ref{SUBSECbilinear}, we have a consistent truncation along $\Phi=\bar{\Phi}$ provided the function $g$ is symmetric under the exchange $\Phi \leftrightarrow \bar{\Phi}$. In the following, we will then assume $g$ to be a real and symmetric function of its arguments.

Supersymmetry is still broken purely along the $S$-direction, with the F-terms equal to
\begin{equation}\label{Ftermslinear}
D_\Phi W=0\, \qquad D_S W = M + g(\Phi,\bar{\Phi})W_0\,.
\end{equation}
Note that the term in $S$ now receives a $\Phi$-dependent correction.

The scalar potential of this model now involves derivatives of $g$, and is given by
\begin{equation}\label{potlinear}
V(\Phi) = - 3 W_0 ^2 + \frac{\left[M + W_0\ g(\Phi,\Phi)\right]^2}{1 - g'(\Phi,\Phi)^2} ,
\end{equation}
along the inflationary trajectory $\Phi=\bar{\Phi}$.

Contrary to the models with a correction coupling to $S \overline{S}$, the task of finding the form of $g$ which yields the desired inflationary potential $V$ now requires solving a non-linear differential equation.  This makes constructing models with low-scale supersymmetry breaking, such as the example \eqref{m2phi2lowscale}, an intractable problem.

However, we can make some progress. In particular, in the regime
\be
|g| \ll 1 \;\; , \;\; |g'| \ll 1 ,
\ee
the potential can be expanded perturbatively in $g$ and $g'$, as follows
\be\label{potlinearsmall}
V = (M^2 - 3 W_0^2) + 2 M W_0\ g + \mathcal{O}(g^2, g'^2)\,,
\ee
which is similar to the expansion \eqref{potbilinearsmall}. Therefore, the same quadratic model Eq.~\eqref{quadraticpotential} can be constructed here via the choice
\be
g(\Phi,\bar{\Phi}) = \frac{m^2}{2MW_0} \Phi \bar{\Phi}.
\ee
As in \eqref{m2phi2perturbative}, the normalization of the inflationary potential in conjunction with the requirement that $|g|\ll1$ forces $M$ and $W_0$  to unobservably large values, corresponding to high-scale supersymmetry breaking.

\pagebreak
\section{$\alpha$-Attractors from K\"ahler corrections to hyperbolic geometry}\label{SEC.HYP}

In the previous section we have considered the case of K\"{a}hler corrections which mix a nilpotent superfield $S$ and a chiral one $\Phi$, where the latter spans a flat internal manifold (i.e. zero \Kahler curvature).  However, typical \Kahler potentials arising from string theory compactifications have often a logarithmic dependence on the moduli and describe a {\it hyperbolic geometry} (see \cite{Hyp} for an analysis of its properties in relation with the physical implications). When the latter is expressed in terms of half-plane variables, the presence of an $\overline{\rm D3}$ brane and a bulk field may be described by the following \Kahler potential:
\be
\label{logk}
K = - 3 \alpha \log \left( \Phi + \bar{\Phi}\right) + S \bar{S}\,,
\ee
where the parameter $\alpha$ controls the value of curvature of the internal field-space, given by $R_K = -2 /3\alpha$.

One can  make the inversion and rescaling symmetries of this \Kahler potential explicit by performing a \Kahler transformation (which leaves the physics invariant) and obtain \cite{Hyp2}
\be\label{logkshift}
K = - 3 \alpha \log \left(\frac{\Phi + \bar{\Phi}}{2|\Phi|} \right) + S \bar{S}\,.
\ee
The latter can be regarded as the {\it curved} analogue of the flat and shift-symmetric \Kahler potential \eqref{KWnil}. It indeed vanishes at $\Phi=\bar{\Phi}$ and $S=0$ and it is again explicitly symmetric with respect to a shift of the canonically normalized field 
\be\label{canonicalfield}
\varphi= \pm\sqrt{\frac{3\alpha}{2}} \log \Phi\,.
\ee
The \Kahler potential \eqref{logkshift}  then implies non-trivial kinetic terms of the field $\Phi$, such as
\be
K_{\Phi \bar{\Phi}}\ \partial \Phi \partial \bar{\Phi}= \frac{3\alpha}{\left(\Phi+\bar{\Phi}\right)^2}\  \partial \Phi \partial \bar{\Phi}\,,
\ee
thus inducing a boundary in moduli space, placed at both $\Phi\rightarrow 0$ and $\Phi\rightarrow \infty$ (note the symmetry under $\Phi\leftrightarrow 1/\Phi$).

When the field $\Phi$ moves away from this boundary, in the direction $\Phi=\bar{\Phi}$, the inflationary implications are very peculiar as they generically lead to a scalar potential which is an exponential deviation from a dS phase such as
\be\label{plateauinflation}
V=V_0 + V_1\exp\left({-\sqrt{2/3\alpha}\ \varphi}\right) + \ldots\,,
\ee
when expanded at large values of the canonical field $\varphi$. This yields universal cosmological predictions in excellent agreement with the latest observational data \cite{CosmologicalData}.

Some working examples of this phenomenon were already found in \cite{HypExamples}. However, the general framework with a varying \Kahler curvature in terms of the parameter $\alpha$ was developed by \cite{alphaattractors} and the corresponding family of models has been dubbed `$\alpha$-attractors'. Further studies have clarified that the attractor nature is simply a peculiar feature of the \Kahler geometry of the sole inflaton sector, independently of the SUSY breaking directions and with a certain special resistance to the other fields involved \cite{OtherFields}. It can indeed be  realized by means of a single-superfield setup \cite{alphaattractorssingle} (see also \cite{GL}, in the case of flat geometry). The case where the bulk field $\Phi$ is coupled to a nilpotent sector $S$, via a \Kahler potential equal to \eqref{logk} or \eqref{logkshift}, has been investigated by \cite{NilpotentInflation2}.

In all the works cited above, the inflationary attractor dynamics arises due to a $\Phi$-dependent superpotential $W=W(\Phi,S)$. In the \Kahler frame defined by Eq.~\eqref{logkshift}, it becomes manifest that such a $W$ simply breaks the original scale-symmetry in $\Phi$ of the system (corresponding to a shift-symmetry in $\varphi$) thus generating non-trivial cosmological dynamics. Conversely, a $\Phi$-independent $W$ produces again a pure de Sitter phase such as the one given by Eq.~\eqref{potdS}.

One can then proceed in analogy to the previous Sec.~\ref{SEC.FLAT} by including mixing terms of the form
\be
\delta K = f(\Phi,\bar{\Phi}) S \bar{S} + g(\Phi,\bar{\Phi}) (S +\bar{S})\,,
\ee
in the \Kahler potential \eqref{logkshift}, while keeping a superpotential just dependent on $S$, such as $W=W_0 + M S$. The resulting situation strikingly resembles the flat one and we find simply the same formulas in terms of the geometric variable $\Phi$. Therefore, SUSY is broken just in the $S$ direction as given by Eq.~\eqref{Fterms}, in the case of bilinear nilpotent corrections, and as given by Eq.~\eqref{Ftermslinear}, in the case of terms linear in $S$. In these two cases, the scalar potential takes the form \eqref{potbilinear} and \eqref{potlinear}, respectively. The stability conditions results to be identical to the flat case as well.

There are, however, some important differences  with respect to the flat case, which are worth highlighting:
\begin{itemize}

\item Inflation happens  around the boundary of moduli space at $\Phi\rightarrow 0$ (or $\Phi\rightarrow \infty$). This implies that any polynomial \Kahler correction in $\Phi$ (or in $1/\Phi$, if we expand around infinity), e.g. such as
\be
f= \textstyle\sum_{n=1}^\infty f_n |\Phi|^n\,,\quad g=\sum_{n=1}^\infty g_n |\Phi|^n\,,
\ee
with $f_n$ and $g_n$ some coefficients, is naturally small during inflation (i.e. $|f|\ll1$ and $|g|\ll1$, for bilinear and linear corrections respectively). In the inflationary regime, one can then consider the expansions \eqref{potbilinearsmall} and \eqref{potlinearsmall}.

\item Unlike the flat case, the geometric variable $\Phi$ has non-trivial kinetic terms, being related to the canonical field $\varphi$ by means of Eq.~\eqref{canonicalfield}. This implies that any pertubative \Kahler correction in $\Phi$ will naturally turn into exponential terms in the scalar potential $V(\varphi)$, thus easily allowing for plateau inflation as given by Eq.~\eqref{plateauinflation}.

\item In the case of small \Kahler corrections, one realizes an exponential fall-off such as Eq.~\eqref{plateauinflation} with
\be\label{plateauenergy}
V_0 = M^2-3W_0^2\,,
\ee
and $V_1= M^2$, for bilinear nilpotent terms (see Eq.~\eqref{potbilinearsmall}), and $V_1= 2 M W_0$, for linear nilpotent terms (see  Eq.~\eqref{potlinearsmall}).

\item The pure nilpotent acceleration phase, equal to \eqref{plateauenergy}, thus serves as the Hubble inflationary energy rather than the CC (see e.g. Eq.~\eqref{quadraticpotential}), whereas the perturbative \Kahler mixing terms induce the inflationary slope. Qualitatively, an $\overline{\rm D3}$ brane provides the primordial acceleration which then gets a dependence on $\Phi$, due to the interaction with the bulk geometry. 

\end{itemize}

On the other hand, similar to the case of flat \Kahler geometry, small corrections correspond to very high SUSY breaking scale, which is order Hubble or higher (the compensation between $M$ and the gravitino mass $m_{3/2}=W_0$ determines indeed the inflationary plateau as given by Eq.~\eqref{plateauenergy}). Nevertheless, one can still obtain a desirable low value of the gravitino mass as this results again to be decoupled from the $M$. The contribution of the \Kahler corrections to the SUSY breaking scale might indeed become important at the minimum of the potential (see Eq.\eqref{CC} for bilinear corrections).

Finally, also in the hyperbolic case, the model allows for a residual cosmological constant. This is given by the finite contributions of the \Kahler correction terms at the minimum of the potential, which can be placed at $\Phi=1$ (provided we impose some conditions on the first derivatives of the functions $f$ and $g$). 

\vspace{-2mm}
\subsubsection*{Examples}

We conclude this section with some concrete examples by focusing just on bilinear nilpotent corrections ($g=0$). 

One can easily obtain a simple $\alpha$-attractor model by considering 
\be\label{Fhyperbolicexample}
F(\Phi)= 1+ \sum_{n=1}^\infty \frac{c_n}{M^2} \Phi^n\,,
\ee
which is still related to $f$ by means of Eq.~\eqref{functionF}. Therefore, during inflation (at $\Phi\simeq0$), $F\simeq1$ which corresponds to very small \Kahler corrections $f$. Note that, for some choices of the coefficients $c_n$, Eq.~\eqref{Fhyperbolicexample} has provided quadratic inflation in the flat case (see Eq.~\eqref{Fquadraticinflation}). However, once we assume hyperbolic \Kahler geometry for the bulk field $\Phi$, the corresponding scalar potential reads
\be
V= M^2 - 3W_0^2 + \sum_{n=1}^\infty c_n e^{-\sqrt{\tfrac{2n^2}{3\alpha}}\varphi}\,,
\ee
in terms of the canonical inflaton $\varphi$ and obtained by means of Eq.~\eqref{potbilinear}. The minimum of the potential can be set at $\varphi=0$ (i.e. $\Phi=1$), provided $\partial_\Phi F|_{\Phi=1} =0$, that is
\be
\sum_{n=1}^\infty n\ c_n =0\,.
\ee

One can then control the residual cosmological constant of this model, at the vacuum of the potential, by tuning the several contributions, which add to
\be
\Lambda=V(\varphi=0)= M^2 - 3W_0^2 + \sum_{n=1}^\infty c_n\,.
\ee
Although the magnitude of $M$ is order Hubble (or higher), the gravitino mass $m_{3/2}=W_0$ can be still tuned to phenomenologically desiderable values (e.g. order TeV).

This framework allows for remarkable phenomenological flexibility and one can reproduce several other known models of inflation. Another example is given by the so-called `E-model' \cite{Kallosh:2015lwa}, defined by the potential
\begin{equation}
\label{Emodel}
V= V_0 \left(1 - e^{-\sqrt{\frac{2}{3\alpha}}\varphi} \right)^{2n}\,,
\end{equation}
which for $n=1$ and $\alpha=1$ returns the original Starobinsky model of inflation \cite{Starobinsky:1980te}. This is realized via the choice 
\begin{equation}
F(\Phi) =  \frac{V_0}{M^2}\left(1 - \Phi \right)^{2n} + F_0,
\end{equation}
where the constant $F_0$ can be tuned in order to change the residual CC ($F_0=3W_0^2/M^2$ in the case of Minkowski vacuum).

\section{Discussion}\label{SEC.Discussion}

In this work we have developed models of inflation in supergravity where inflation is driven by terms in the K\"{a}hler potential which mix the inflaton field with a nilpotent superfield, even in the absence of a superpotential for the inflationary sector. The physical situation one would have in mind is given by an anti-D3 brane interacting with a bulk geometry.  We have studied the effects of these additional terms when the internal geometry of the bulk field is either flat or hyperbolic, and found that this generically allows for inflation that exits to de Sitter space. The outcome is a scenario which allows for flexible phenomenology in terms of inflation, dark energy and supersymmetry breaking. 

A general feature of these models is that SUSY is broken purely in the direction of the nilpotent superfield\footnote{This simplifies the description of the fermionic sector, whose non-linear terms disappear from the supegravity action, as already pointed out in the last reference of \cite{NilpotentInflation}.} $S$ for the entire cosmological evolution, thus providing alone the necessary acceleration for inflation and the residual CC. Interestingly, the non-trivial \Kahler corrections (which cannot be gauged away by a \Kahler transformation) become the fundamental ingredient in order to have controllable level of supersymmetry breaking and dark energy at the vacuum of the potential (see the example defined by Eq.~\eqref{Fhyperbolicexample}).

The regime of small \Kahler corrections is definitively important as one would expect these terms arising as subleading dynamical effects. The case of hyperbolic geometry is particularly interesting as perturbative \Kahler corrections in the inflaton $\Phi$ are naturally small (as inflation happens at $\Phi\simeq0$) and the consequent cosmological dynamics is an exponential deviation from a dS plateau at the Hubble scale. The physical picture is that of an anti-D3 brane, responsible for the inflationary acceleration, whose interaction with the bulk geometry induces the typical behaviour of $\alpha$-attractors. 

While we have explicitly studied the case of a $\Phi$-independent superpotential, where the inflationary dynamics is purely \Kahler driven, one may wonder what happens if the \Kahler corrections considered here are incorporated into a model of inflation that is driven by the superpotential. In this case, $W$ acquires a dependence on the inflaton, such as 
\be
W(\Phi,S)= A(\Phi) + B(\Phi) S\,.
\ee
One can prove that, in the case of hyperbolic \Kahler geometry defined by Eq.~\eqref{logkshift}, any Taylor expansion of the functions $A$ and $B$ in the geometric field $\Phi$ will contribute to the scalar potential $V(\varphi)$ with exponential terms, thus preserving the typical attractor behaviour \eqref{plateauinflation}. The situation is different in the case of flat \Kahler geometry, as one generically needs a higher amount of fine-tuning in order to preserve the original superpotential-driven model of inflation. As a clear example of this circumstance, the famous model \cite{Kawasaki:2000yn} of quadratic inflation, defined by $W=m \Phi S$, will be immediately spoiled by any generic polynomial \Kahler correction in $\Phi$.

Another interesting avenue of research would be to investigate the consequences of the modified K\"{a}hler potentials studied here for the fermions, both at early and late times. Such K\"{a}hler potentials will lead to derivative interactions of the inflaton with the fermion bilinear, e.g.
\begin{equation}
\partial \Phi \, \partial \left( \psi \psi \right)
\end{equation}
which may have important consequences for (p)reheating, or leave an imprint in the spectrum of primordial curvature perturbations.

Finally, we note that there are many possible generalizations of the models presented here, similar to the series of developments of superpotential-driven models of inflation considered in \cite{NilpotentInflation}. It would be interesting to understand the extent to which the same is possible for K\"{a}hler potential driven models of inflation.

\section*{Acknowledgements}

We thank Andrei Linde and Timm Wrase for very helpful comments on a draft of this work. Further, we gratefully acknowledge stimulating discussions on related topics with Robert Brandenberger, Wilfried Buchm\"uller, Jim Cline, Keshav Dasgupta, Lucien Heurtier and Alexander Westphal. EM is supported by the The Natural Sciences and Engineering Research Council of Canada (NSERC) via a PGS D fellowship. MS acknowledges financial support by the collaborative research center SFB 676 and by `The Foundation Blanceflor Boncompagni Ludovisi, n\'ee Bildt'.

\end{document}